\documentclass{article}
\usepackage[utf8]{inputenc}
\usepackage{newunicodechar}
\newunicodechar{́}{\'{}}
\usepackage[T1]{fontenc}
\usepackage{parskip}
\usepackage{amsmath,amssymb}
\usepackage{amsthm}
\usepackage{siunitx}
\usepackage{braket}
\usepackage{pdfpages}
\usepackage[english]{babel}
\usepackage{csquotes}
\usepackage[dvipsnames]{xcolor}

\usepackage[a4paper,margin=1in]{geometry}
\usepackage{tabularx}
\usepackage[shortlabels]{enumitem}
\usepackage{graphicx}
\usepackage{titling}
\usepackage[backend=biber,style=apa]{biblatex}
\usepackage[colorlinks=true, urlcolor=blue, linkcolor=black]{hyperref}
\graphicspath{ {./afbeeldingen/} }

\title{Foundations for a Lens of Quantum Governance}
\begin{document}
\begin{titlepage}
\centering

{\LARGE \textbf{Foundations for a Lens of Quantum Governance} \par} 

\vspace{1.2cm}

{\large
Shira K. Huizing Lemstra\(^{1}\) \\
Tzula B. Propp\(^{2,}\)\(^{3}\)
\par}

\vspace{1.5cm}

{\small
\(^{1}\)  
Faculty of Architecture and the Built Environment, Delft University or Technology, \\ 
Building 8, Julianalaan 134, 2628 BL Delft.  
\\Email: \texttt{s.k.huizinglemstra@tudelft.nl}
\par}

{\small
\(^{2}\)  QuTech, Kavli Institute of Nanoscience, Delft University of Technology, \\ Lorentzweg 1, 2628 CJ Delft, Netherlands}

{\small
\(^{3}\)Leiden Institute of Physics (LION), Leiden University, \\Niels Bohrweg 2, 2333 CA Leiden
\\Email: \texttt{propp@physics.leidenuniv.nl}
\par}

\vfill

\end{titlepage}

\section*{\centering Abstract}
\addcontentsline{toc}{section}{Abstract}
    The second quantum revolution is a revolution of information. Advances in quantum computation, communication, and sensing promise capabilities that extend beyond those of existing digital technologies, including new possibilities for secure communication, distributed computation, and precision measurement. Although much of the discussion surrounding these developments focuses on their technical potential, their implementation also raises questions about how emerging quantum infrastructures will be organised, accessed, and governed. We argue that the significance of quantum technologies for governance lies not in their capacity to fundamentally transform political principles, but in how their particular capabilities and infrastructural requirements may influence existing relationships of authority, responsibility, legitimacy, and power.

    In this paper, we develop the foundations of a quantum governance lens that brings established perspectives on governance and sociotechnical systems into dialogue with quantum information science. Rather than proposing an entirely new model of governance, this lens provides an analytical framework for examining how technical design, institutional arrangements, and infrastructural dependencies interact. Drawing on scholarship concerning administrative legibility, infrastructures, sociotechnical imaginaries, interscalar relationships, and institutional accountability, we identify the importance of examining uncertainty, relational authority, the allocation of responsibility, institutional responsiveness, privacy, and access to specialised quantum resources.

    We illustrate the relevance of this lens through five applications: quantum computation, quantum key distribution, distributed consensus through Quantum Byzantine Agreement, private delegated computation via blind quantum computation, and dual-use\footnote{That is, having both civilian and military applications.} quantum sensing. These cases demonstrate how particular technical capabilities and requirements may affect the organisation of access, trust, oversight, and accountability. They also highlight opportunities for secure communication, privacy-preserving computation, and new forms of coordination, while showing that technical guarantees alone cannot resolve questions of institutional responsibility or the distribution of infrastructural power.

    We conclude that the governance challenges associated with quantum technologies are not necessarily unique to quantum systems, nor do they establish the inadequacy of existing governance frameworks. Rather, emerging quantum infrastructures provide a context in which familiar questions of legitimacy, accountability, and equity must be examined in relation to specific technological capabilities and institutional choices. These concerns are already visible in the development of other digital technologies, particularly artificial intelligence. By connecting quantum information science with governance scholarship, this paper offers a foundation for interdisciplinary research into how emerging technologies can be developed and governed in ways that support accountability, privacy, cooperation, and equitable access.

\newpage

\section{Introduction}

The internet has become so deeply embedded in modern life that its social, political, and economic significance is often taken for granted; the internet, and by extension, the technological infrastructure that underpins it, structures communication, world markets, governance, cultural imagination, diplomacy, and the day-to-day lived experiences of billions of people. 

As Peter P. Rohde and colleagues (2025) discuss in their interdisciplinary roadmap for the development of a global quantum internet, the coming decades may witness significant advances in quantum networking technologies, with potentially far-reaching societal consequences. A quantum internet refers to a network of interconnected quantum devices, including quantum computers, quantum sensors, repeater stations, and client devices, that can distribute quantum states and entanglement between nodes. While technically challenging to build, a quantum internet could enable tasks that are inefficient or impossible using classical networks and make distributed quantum computation accessible to end users. Whereas the first quantum revolution enabled technologies grounded in our understanding of quantum phenomena, including transistors and lasers, the second quantum revolution is characterised by the capacity to manipulate and harness individual quantum systems for information processing, communication, and sensing (Dowling \& Milburn, 2002).

\begin{figure}[b!]
    \centering
    \includegraphics[width=0.45\textwidth]{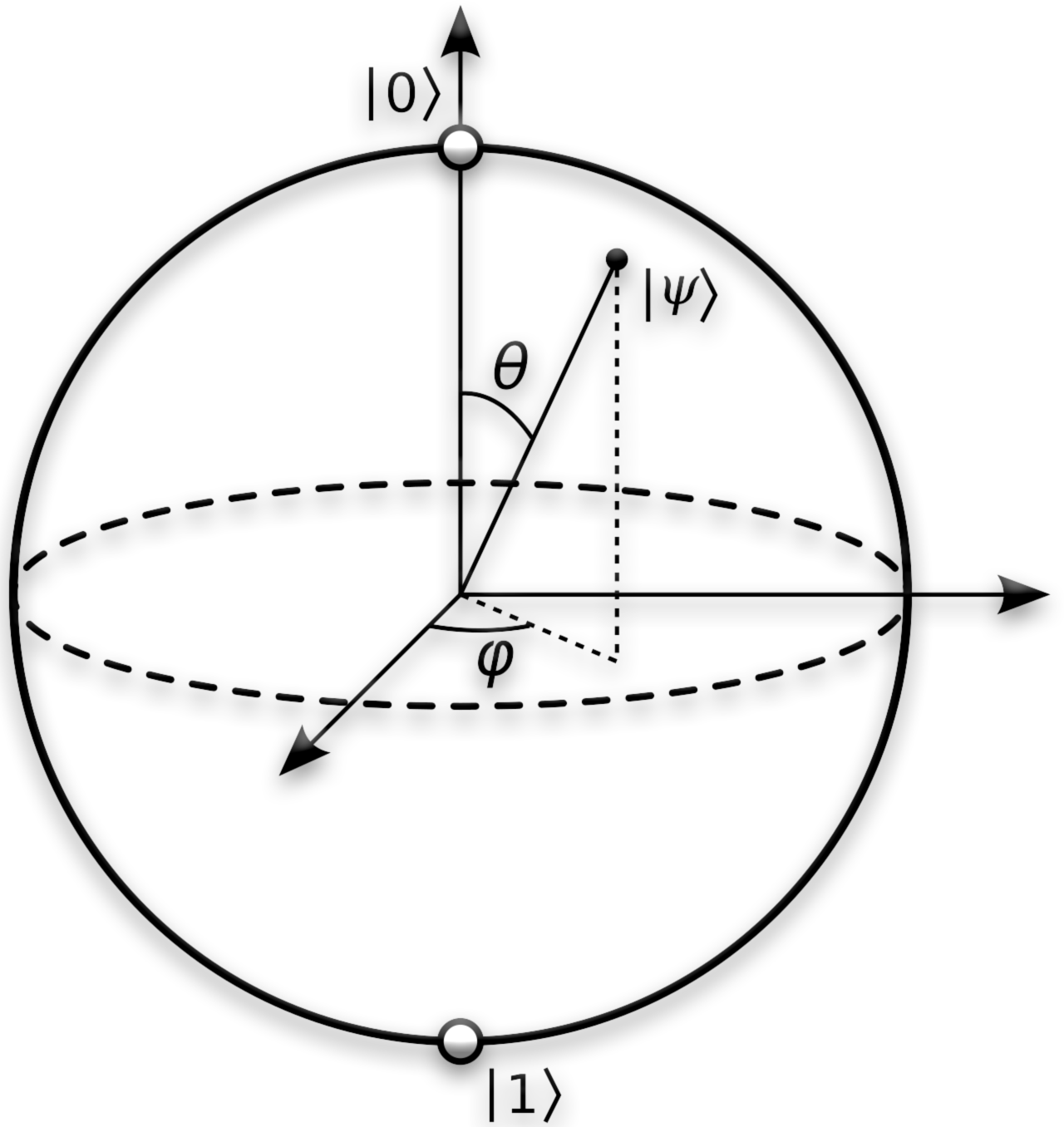}
    \caption{
       The state $\ket{\psi}$ of a qubit is represented with a point on the Bloch sphere. Here, the coefficients in the superposition are parameterized in terms of longitudinal and latitude, such that $\ket{\psi} = {\rm cos}(\frac{\theta}{2}) \ket{0} + e^{i\phi} {\rm sin}(\frac{\theta}{2}) \ket{1}$ with $\theta\in\{0,\pi\},\phi\in\{0,2\pi\}.$
    }
    \label{fig:bloch-sphere}
\end{figure}

The power of a quantum protocol, either on a computer or a network, lies in its ability to harness entanglement and superposition of quantum bits, or qubits, the fundamental building block of modern quantum information science. One common representation of a qubit is the Bloch sphere, shown in Fig. \ref{fig:bloch-sphere}; here, the qubit state $\ket{\psi}$ is represented by points on the surface of a ball.\footnote{Furthermore, the interior of the Bloch sphere corresponds to statistical mixtures of the states $\ket{0}$ and $\ket{1}$ i.e. states with reduced coherence. This requires going beyond the wavefunction formalism and introducing density matrices. For more information, see Nielsen \& Chuang (2010).} A classical bit of information can take the value $0$ or $1$. A physical system encoding this information can be thought of as a coin, which can be in the states ``heads'' ($H$) or ``tails '' ($T$). Whereas a classical coin can only be $H$ or $T$, or a statistical mixture of the two, a quantum coin can be in an arbitrary superposition state: $\ket{\psi} = a\ket{H} + b\ket{V}$, with $\ket{\psi}$ the standard notation for a complex vector describing the quantum state of the coin (or bit) in Hilbert space (the mathematical space where quantum states live), and where we require $|a|^2 + |b|^2 = 1$ for the conservation of probability and normalization of the quantum state. Here $\ket{H}$ and $\ket{V}$ correspond to when a quantum coin is in the state heads or tails, respectively; that these mutually exclusive outcomes may be added in combination to form a new, permissible quantum state is the essence of the superposition principle of quantum theory. For conceptual and mathematical introductions, see Wong (2022) and Nielsen \& Chuang (2010), respectively. 

By designing protocols that utilise superposition as well as entanglement at the level of hardware, quantum information enables new applications. This includes quantum key distribution (QKD) for unconditionally secure peer-to-peer communication, secure multi-party cloud-based quantum computation and quantum-enhanced artificial intelligence models, and large-scale entanglement distribution enabling new forms of coordination, sometimes referred to as “quantum games” (Wehner, Elkouss, \& Hanson, 2018) building upon the original quantum games literature (Meyer, 1999; Eisert, 1999). 

These developments may have implications beyond their immediate technical applications, particularly for the organisation and governance of digital infrastructures. We understand governance as the institutional, infrastructural, and normative arrangements through which authority, responsibility, and legitimacy are organised in sociotechnical systems. These technological developments may reshape the conditions under which digital authority, trust, accountability, and power are organised. Rohde et al. (2025) examine the economic and political implications of a future quantum internet, including markets for computational resources, taxation, centralised resource allocation, and the geopolitical consequences of unequal access.  They envisage "a marketplace for tradable units of quantum computation, where the underlying asset is time-shares in the global network" (p. 259), while also recognising that dependence on shared computational resources may create political vulnerabilities. Although questions of resource allocation and political power are not unique to quantum technologies, the technical characteristics of quantum information may alter how these relationships develop. 

This raises a broader question of how governance itself may be shaped by the infrastructures it seeks to govern. Rather than assuming that quantum technologies require entirely new political principles, we examine how their properties and applications interact with existing arrangements of authority, responsibility and legitimacy. This approach draws on Scott (1998) and Edwards (2019), whose work higlights how infrastructures and practices of knowledge production shape the possiblities and limitations of governance.  

The recent experience of algorithmic governance underscores this challenge. As Zuboff (2019) argues in her analysis of surveillance capitalism, digital platforms have turned human experience into behavioural surplus: data is used to predict and increasingly to shape behaviour. This shift, “from automating information flows about you to automating you” (Zuboff, 2019, p. 8), reveals how infrastructures can embed forms of power that elude traditional regulatory frameworks, and how the prediction industry has co-created governance with global ramifications without the consent, or even knowledge, of the general public. The anticipated wave of quantum-enhanced AI may further intensify this dynamic (cf. Rohde et al., 2025), as ever more powerful computational tools interact with opaque and probabilistic infrastructures. This parallel suggests that the need for new governance approaches is not limited to quantum technologies but reflects a broader challenge in governing systems whose internal processes and external effects cannot be rendered fully legible. 

Applying these insights to a future quantum internet allows us to examine how governance decisions shape technological infrastructures, and how those infrastructures may, in turn, influence the exercise of authority and responsibility. A future quantum internet could introduce capabilities that differ from those of existing digital networks, including entanglement distribution and networked quantum computation. The development and organisation of these infrastructures may affect how computational resources are accessed, how trust is established, and how authority is exercised. They also raise questions about technological reliability, dependence on specialised networks, and the distribution of access and control. Their governance involves uncertainty about how these technologies will develop, who will control them, and how their benefits and risks will be distributed. We therefore examine whether existing governance frameworks adequately account for these developments, rather than presupposing that they must be replaced.

This raises a central question: how might the development of quantum information infrastructures affect existing arrangements of governance, particularly where technical interdependence and specialised resource requirements introduce new challenges for coordination, accountability, and control? To address this question, we draw on several approaches to governance and sociotechnical systems. We use Scott's (1998) critique of legibility to examine the limits of centralised control, and Edwards' (2003, 2019) work on infrastructures to understand how technical systems shape authority and responsibility. Zuboff's (2019) analysis of surveillance capitalism illustrates the consequences of allowing such infrastructures to develop without sufficient public accountability. We then draw on Jasanoff (2020) to examine how competing visions of the future influence present decisions about quantum technologies. Hecht's (2018) work helps us consider how responsibility operates across different geographical and temporal scales, while Boin, McConnell, and 't Hart (2008) provide a basis for understanding accountability and institutional learning under uncertainty. Together, these perspectives allow us to examine governance as a process shaped by both institutional decisions and the infrastructures through which they operate.

Our contribution is to develop the foundations of a quantum governance lens that brings insights from governance theory into dialogue with the technical characteristics of quantum information infrastructures. This lens provides a way to examine how existing questions of authority, responsibility, trust, and legitimacy may take on different forms as quantum technologies develop. This reframing also has implications for contemporary debates on AI governance, which likewise confront systems that resist reduction to legible and predictable forms. Scholars have shown that AI systems challenge classical governance expectations by relying on opaque, distributed, and often unintelligible forms of decision-making (Ananny \& Crawford, 2018), by converging around high-level ethical principles that provide little practical guidance (Floridi \& Cowls, 2019), and by producing accountability gaps that existing regulatory and professional frameworks struggle to address (Mittelstadt, 2019). In this, it also resonates with Boin et al's emphasis on accountability and learning in post-crisis governance, where legitimacy depends not on claims of omniscience but on the capacity to adapt, reflect, and remain answerable under conditions of disruption (Boin, McConnell, \& ’t Hart, 2008).

This paper thus examines how the technical characteristics and applications of quantum information technologies may interact with existing arrangements of authority, responsibility, and legitimacy. We develop the foundations of a quantum governance lens to analyse these interactions and illustrate its relevance through selected applications of a future quantum internet.

\section{Literature background}
\subsection{A quantum internet as infrastructure}

Rohde et al. (2025) describe a quantum internet as central to the “second quantum revolution”, a development expected to transform global infrastructures much as the classical internet did in the twentieth century (Dowling \& Milburn, 2002). They write that it will be “a distributed quantum network-of-networks, linking quantum computers, sensors and communication devices into a global system” (p. 239). Among the applications they highlight are “quantum key distribution for unconditionally secure communication” (p. 244), “cloud-based quantum computation” (p. 243), “blind quantum computation, which allows a user to perform a computation on a remote quantum server without revealing either the data or the algorithm” (p. 244), and the large-scale “distribution of entanglement … as the fundamental resource enabling coordination between distant nodes” (pp. 246-257), where entanglement refers to non-classical correlations between quantum systems that cannot be reduced to the properties of individual components. This coordination often takes the form of quantum state teleportation, wherein the state of a quantum system is transferred between distant nodes using previously shared entanglement and classical communication. Quantum teleportation transfers an unknown quantum state without creating a second copy of it, consistent with the no-cloning theorem, which prohibits the perfect replication of an arbitrary unknown quantum state (Wootters \& Zurek, 1982). The no-cloning principle also underlies Wiesner's (1983) proposal for unforgeable quantum money.

In its simplest implementation, quantum state teleportation consumes one maximally entangled Bell pair, or ebit, to teleport the quantum state of a qubit across space. While the teleportation of the state is instantaneous, the reconstitution of the state is limited by the classical communication time between locations, recovering consistency with Einstein’s relativity (Bennett et al., 1993).

What makes a quantum communication-and-computation platform capable of implementing quantum protocols? This is captured by the DiVincenzo criteria, which specify the basic requirement for a universal quantum computer as: “a scalable physical system with well-characterised qubits; the ability to initialise the state of the qubits; long relevant decoherence times; a universal set of quantum gates; and the ability to measure specific qubits” (Nielsen \& Chuang, 2010, pp. 186-187). For a quantum internet, two further criteria are necessary: “The ability to interconvert stationary and flying qubits and to faithfully transmit flying qubits between distant nodes” (Nielsen \& Chuang, 2010, pp. 478-479; see also Wong, 2022, pp. 175-176). Here, stationary qubits refer to quantum information stored locally at a network node, whereas flying qubits carry quantum information between nodes, typically through photons. Interconversion between the two enables quantum information to be stored, processed, and transmitted across a network. In summary, a quantum internet depends on quantum processing nodes connected by classical networks, as well as physical distribution of entangled quantum states across space via a co-deployed quantum network, connecting nodes capable of both processing and preserving quantum information. 

Quantum computers are built out of quantum bits or qubits. One simple representation of a qubit is given by the Bloch sphere (see Fig. \ref{fig:bloch-sphere}). Whereas a classical bit can take only two values, 0 or 1––or in an ensemble a probabilistic mixture of the two––the quantum bit can occupy a whole range of pure states denoted by the state vector $\ket{\psi}$. The north and south pole correspond to the logical qubit states $\ket{0}$ and $\ket{1}$, respectively. A measurement of a quantum state requires specifying a basis, i.e.~an axis along the Bloch sphere, with states antipodal on the sphere representing mutually exclusive states; mathematically, these correspond to a vanishing inner product $\braket{1|0} = 0$. Single qubit gates are represented by rotations on the Bloch sphere. Because a qubit can be in superposition of the logical qubit states $\ket{0}$ and $\ket{1}$, the state space (i.e. Hilbert space) available is larger on a quantum computer than on a classical computer. This is part of the reason for the computational advantage enabled by quantum computation. 

The power of quantum computation also comes from the ability to entangle qubits. Entangling two qubits e.g. in a state $\ket{\Psi} = \frac{1}{\sqrt{2}}
\left(\ket{0}\ket{0} + \ket{1}\ket{1}\right)$ results in non-separability; one cannot give a complete description of either qubit alone. Conceptually, the two-qubit state $\ket{\Psi}$ corresponds to a case where both qubits are guaranteed to be in the same state, corresponding to the values $0$ or $1$, but the two possible states with the same values for the two qubits ($ \ket{0}\ket{0} $ and $ \ket{1}\ket{1}$) have been added together in superposition. This correlation between states persists regardless of what basis is chosen on the Bloch sphere to measure both qubits. If different bases are chosen, one can utilize the correlations to e.g. show violations of Bell’s theorem (Storz et al., 2023). In practice, entanglement is a vital resource in quantum networks; consuming one entangled pair of qubits allows for the teleportation of an unknown qubit state as described above. However, it is hard to entangle qubits on remote quantum servers, hence the necessity of quantum repeaters to distribute entanglement to distant nodes (Azuma et al., 2023). 

However, not just any entanglement or superposition enables beyond-classical computational ability; a key part of quantum computation is the reliance on so-called magic states (Bravyi, 2005). Certain quantum circuits, including those restricted to stabiliser operations, can be simulated efficiently on classical computers despite involving superposition and entanglement. Magic states provide an additional resource that, when combined with these operations, enables universal quantum computation and can make classical simulation substantially more difficult (Bravyi, 2005).

A key element in quantum computing is the use of a universal gate set which always includes at least one entangling gate that acts between qubits. As Nielsen \& Chuang (2010) explain: “an entangling two-qubit gate together with arbitrary single-qubit gates is universal for quantum computation” (p. 185). Entanglement, they discuss in their canonical textbook, “creates correlations that cannot be expressed as properties of the individual subsystems” (p. 97) and, because of the probabilistic nature of measurement, “renders impossible the prediction of outcomes with certainty” (p. 89). Measurement-based quantum computation reduces the entirety of quantum computation protocols to adaptive measurements made of highly entangled graph states (Briegel et al., 2009). This means that a critical capacity of a quantum network arises, not from the properties of its discrete components, but rather from intrinsically holistic, contextual properties spread across the network.

Quantum information is fragile (Terhal, 2013). Interactions with environments degrade the quantum information stored in qubits: a process called decoherence (Zurek, 1991). This fragility is why the first quantum revolution which was based on incoherent interactions (which are necessary e.g. for computers and cell phones) happened almost 100 years before the second ongoing quantum revolution, which is based on the coherent manipulation of quantum information and construction of quantum computers. One breakthrough was in 1994 when Peter Shor discovered the first quantum error correcting code (Shor, 1994), and now decoherence-resistant quantum computation is being developed, with much progress underway (Arute et al., 2019). 

Entanglement in general is a quintessentially quantum holistic property, as it arises from the relation between systems rather than their isolated status. It also renders entangled subelements of a system contextual; a complete description of the system can only be given on the level of the whole and not of the parts. For the development of a quantum governance lens, the relevance of entanglement lies in the technical dependencies it can create between networked systems, and in the questions these dependencies raise about coordination, access, and responsibility.

In their discussion of the impact of next-generation quantum technology on governance, Rohde et al. (2025) present these innovations as socially transformative. They emphasise that the distributed nature of quantum networking changes not only how computation is performed, but also how it is accessed and owned. They propose “a marketplace for tradable units of quantum computation, where the underlying asset is time-shares in the global network. We refer to this as a quantum stock market – a marketplace subject to ordinary supply and demand” (p. 259). They further suggest that “the qubit would be ripe for the application of a consumer-driven tax” (p. 271) and at points envisage that global resources will be “overseen by a trusted central authority, congruent with our efficient-market assumptions” (p. 258). From this we see that the authors expect that the governance of a quantum internet cannot be separated from economic and political stakes, and that access to quantum computing is expected to be organised through market-based allocation of distributed computational resources rather than by physical ownership of single physical machines. 

Importantly, a universal quantum computer can efficiently perform tasks such as factoring large integers using Shor’s algorithm, that are intractable for any known classical computer (Nielsen \& Chuang, 2010, pp. 221–222). As Nielsen and Chuang (2010) stress, “Shor’s factoring algorithm runs in polynomial time on a quantum computer, thereby undermining the security of widely used public-key cryptosystems” (pp. 221–222). This exponential speedup means that large-scale quantum computation threatens to upend the assumptions about adversarial computational limits that currently underpin classical digital security, including the security of online banking, credit-card transactions, and encrypted communications. 

Indeed, NIST warns that “large-scale quantum computers will be able to break the public-key cryptosystems currently in widespread use” (NIST, 2016). Today, most digital payments, including credit-card transactions, online banking, and e-commerce, depend on public-key encryption. When a browser sends payment information to a website, it uses RSA or elliptic-curve cryptography to establish a secure channel. The security of these protocols rests entirely on the assumption that no adversary has the computational power to factor large integers or compute discrete logarithms. A universal quantum computer running Shor’s algorithm breaks both problems efficiently. As a result, an attacker with access to a sufficiently large quantum computer could decrypt intercepted payment traffic, impersonate merchants, forge digital signatures, or read stored financial records protected by these schemes. Because nearly all global payment infrastructures rely on these assumptions, the arrival of large-scale quantum computers represents a direct threat to the confidentiality and integrity of everyday financial transactions. Thus, universal quantum computers can break much of classical encryption, and although there is extensive research into alternatives through post-quantum cryptography (Bernstein, Buchmann, \& Dahmen, 2009), there are still no guarantees that any proposed scheme will be fully resistant to future quantum attacks. In the era of information, the concentration of quantum resources carries potentially unprecedented geopolitical and economic leverage.

It is worth pausing to note that the modern AI revolution has been enabled by the development of the graphical processing unit (GPU), a physical technology. Similarly, quantum computers act as quantum processing units (QPUs), which will enable new quantum algorithms. Indeed, the implications of this shift in technological capacity extend to the domain of artificial intelligence. Work on quantum machine learning suggests that quantum computers can in some (but not all) cases offer provable speed-ups over classical algorithms (Biamonte et al., 2017, pp. 195–196; Huang, Kueng, \& Preskill, 2021, pp. 2-3; see also Tang, 2019, for classical algorithms that match certain claimed quantum advantages). As the ongoing AI revolution already challenges existing regulatory frameworks, the integration of quantum computation into AI development and deployment is likely to amplify questions of access, accountability, and global inequality (Ananny \& Crawford, 2016).

At the same time, specific applications foreshadow new forms of trust and verification. Quantum key distribution (QKD) promises “unconditionally secure communication … with the feature that any eavesdropping attempt is physically detectable” (Rohde et al., 2025, p. 244), yet whoever controls the infrastructure decides who can benefit from this security. Blind quantum computation (BQC) offers privacy to clients: “the server learns nothing about the data or the algorithm being computed” (Rohde et al., 2025, p. 244). Further modifications for verified BQC, even in the presence of noise, place the ability to verify the blindness of delegated computation squarely in the hands of end users (Leichtle, Music, Kashefi, \& Ollivier, 2021; Gheorghiu, Hoban, \& Kashefi, 2024), even when those end users do not themselves possess quantum computers. Recent work on photonic-client architectures shows that such verification can be achieved with extremely limited quantum hardware on the client side (van Dam et al., 2024, ). This enables new modalities of trust and privacy, and yet simultaneously raises new questions of jurisdiction and liability when computation crosses borders. There are also intersections of AI research and quantum cryptographic research in emerging work on blind quantum machine learning (Fleur, 2025).  While in Vermaas (2025) it is pointed out that utilizing these protocols still requires trust of the underlying technology, it does not require users to do so blindly; verifiable blind quantum computation allows end-users to verify the performance---as well as the capabilities---of the underlying servers; since a quantum server cannot tell a verification round from a computation round, honest behavior is distinguishable from dishonesty. And although the math of quantum theory is hard, it is not opaque; it is a straightforward calculation to show that e.g. entanglement can be used to generate shared random secret keys. A benefit of a science-based approach to privacy is precisely that anyone can, in principle, replicate the results upon which the theories of its functions rest.

Rohde et al. (2025) also point to the broader economic and geopolitical stakes. They note that “the qubit would be ripe for the application of a consumer-driven tax” (p. 271), highlighting that computational power itself is envisioned as a taxable resource. They further suggest that the emergence of a “quantum stock market … subject to ordinary supply and demand” (p. 259) could allow global players to trade and speculate on computational time-shares in network. While these proposals are presented as extensions of familiar market tools, they imply that those who control access to qubit resources - whether states, corporations or hybrid alliances of both - will wield disproportionate leverage over economic activity, security and scientific research.

A quantum internet is thus not only a technical innovation but a contested sociotechnical arena in which political, economic, and social interests intersect and, possibly, clash. The allocation of scarce quantum resources, the design of taxation and pricing models, and the concentration of control in a few actors will shape the future distribution of power and influence on both domestic and international scales. We emphasise that governance in this context must not be understood as an after-the-fact regulatory layer applied to a neutral technical platform. Instead, it should be recognised as co-constituted with the very infrastructures of a quantum internet itself, making the network not merely a technical breakthrough but a site of ongoing struggle over authority, legitimacy, and responsibility.

\subsection{Governance and the limits of legibility}

Scott’s (1998) critique of high modernism highlights the risks of assuming that societies can be rendered fully legible through simplification. Infrastructural projects built on the premise of centralised control, whether in urban planning, agriculture, or language standardisation, have historically often collapsed under the weight of complexity, producing unintended consequences when local knowledge was disregarded. Though the science it is based on is foundational (physics), a quantum internet poses a parallel challenge. While quantum infrastructures share many governance challenges with existing digital systems, their reliance on specialised hardware, entanglement distribution, and the preservation of fragile quantum states may introduce additional technical dependencies that complicate coordination, oversight, and accountability. From Scott's (1998) perspective, the governance challenge lies not in quantum indeterminacy itself, but in whether attempts to simplify these infrastructures for administrative purposes adequately account for their technical dependencies, institutional complexity, and the knowledge required to oversee them. Furthermore, even indeterminism-friendly legibility may not be achievable in the short term given the low scientific literacy around quantum science around the world (Nita, Mazzoli Smith, Chancellor, \& Cramman, 2021). The leaders of our current governments utilize AI without being experts in it themselves, and we may soon find ourselves in a similar situation with quantum technology. Read alongside Rohde et al. (2025), Scott's work invites examination of whether centralised approaches to governing quantum infrastructures adequately reflect their technical and institutional complexity, particularly where oversight depends on specialised expertise.

\subsection{Infrastructures as architectures of power}

Edwards (2003) characterises infrastructures as the hidden architecture of modernity: embedded, standardised, and largely invisible until breakdown. They organise time, space, and force, stabilising authority while simultaneously constraining possibilities for action. In later work, Edwards (2019) shows how knowledge infrastructures act as both memory and battleground, structuring legitimacy and serving as sites of contestation. 

These insights apply directly to a quantum internet. As Rohde et al. (2025) note, questions of allocation, taxation, and leverage will be inseparable from the technical protocols and architectures of the network, which are themselves largely determined by those who design and build it. Governance will not simply oversee these infrastructures; it will be constituted through them. Authority and responsibility will be distributed across human and non-human actors, embedded in the operation of distributed quantum  protocols, cryptographic systems, and marketplace designs.

\begin{figure}[htbp]
    \centering
    \includegraphics[width=0.8\textwidth]{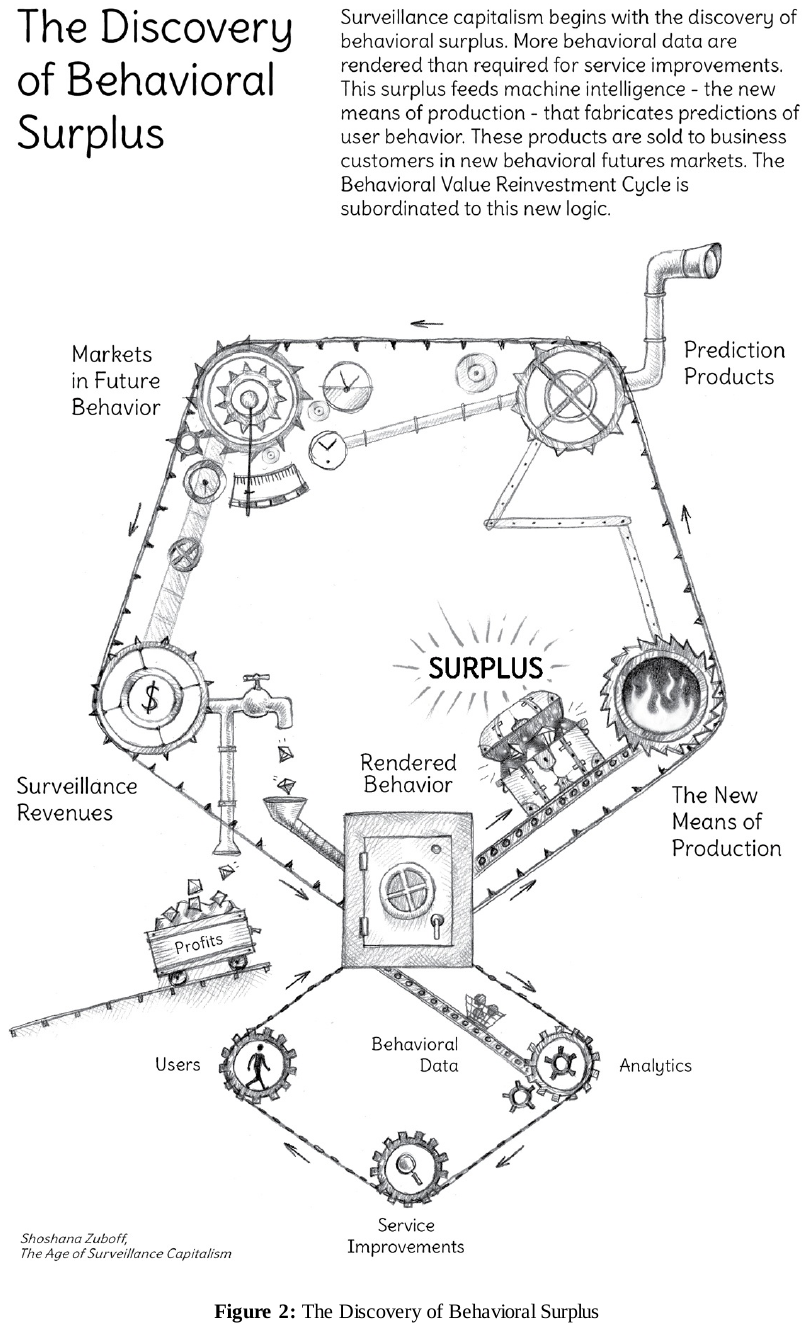}
    \caption{Infrastructures of behavioural surplus in surveillance capitalism; as users interact with e.g. a search engine to recieve results, metadata (time of search, browser cookies) are harvested to generate behavioral predictions and modifications; only a small fraction of the behavioral data is used to improve the search results for the user's benefit. On a quantum internet, blind quantum implementations of search engines, machine learners, recommendation systems, and artificial intelligences enable users to gain access to their requested service without the leaking of behavioral surplus. This figure was reproduced with permission by the author from The Age of Surveillance Capitalism: The Fight for a Human Future at the New Frontier of Power by Shoshana Zuboff. }
    \label{fig:zuboff}
\end{figure}

Fig.~\ref{fig:zuboff} visualises how infrastructures encode asymmetries of power by capturing private experience and converting it into predictive and control capacities. As Zuboff argues, surveillance capitalism extracts “behavioural surplus … the raw material for a new means of production” (2019, p. 93), turning human activity into data flows that can be used not only to predict but increasingly to shape behaviour. This dynamic illustrates that the distribution of power is inseparable from infrastructural design.

The relevance to quantum governance is twofold. First, quantum infrastructures could intensify such dynamics: their opacity and nonlocal correlations may further obscure how data is processed and decisions are shaped, evading legible oversight. Second, certain quantum protocols also offer possibilities for resisting them. BQC allows a client to delegate a computation to a remote, untrusted quantum server while keeping both the input and the algorithm secret (Rohde et al., 2025, p. 244). In principle, BQC limits the server’s ability to harvest the behavioural surplus on which surveillance capitalism depends, allowing users to access powerful computation without surrendering informational sovereignty. Widespread adoption of such technologies will challenge the current information centralization that modern surveillance capitalism relies upon to function. This could be a way to achieve a sort of universal basic income; some degree of income would be guaranteed by the (voluntary) selling of one’s own data to train AI models, in comparison to the contemporary involuntary extraction of training data.

This juxtaposition highlights that governance is not only shaped by infrastructures but can also, in turn, guide the design and adoption of infrastructures to resist particular patterns of power. A quantum internet thus sharpens rather than erases the need to confront how technical architectures distribute authority, responsibility, and legitimacy, and can be used both in support of and in opposition to the preservation of democracy, which is of especially large import in contemporary time.

\newpage

\subsection{Imaginaries and future-making}

Jasanoff (2020) argues that technologies are guided by sociotechnical imaginaries: “collectively held, institutionally stabilised, and publicly performed visions of desirable futures” (p. 15); in short, collective acts of science fiction that shape what we believe is possible, likely, or even inevitable.  Such imaginaries set the horizons of what counts as progress and whose interests are prioritised. They are not merely stories about the future; they actively guide present-day investments, regulations, and infrastructural choices.

Rohde et al. (2025) implicitly mobilise two contrasting imaginaries of a quantum internet. At several points they cast it as a universal good, describing it as “a global resource for secure communication, collaborative science, and distributed computation” (p. 239) and emphasising its potential to “bring unprecedented security and efficiency to global information networks” (p. 244). Elsewhere, however, they frame it in terms of competition and leverage, presenting the prospect of “a marketplace for tradable units of quantum computation … subject to ordinary supply and demand” (p. 259) and warning that “states that dominate access to entangled resources will command disproportionate economic and geopolitical influence” (p. 272).

These competing imaginaries matter because they push governance in very different directions. A vision of a quantum internet as a universal good encourages policies that focus on shared infrastructures, equitable access, and public accountability. By contrast, an imaginary centred on rivalry and leverage directs attention toward securitisation, competitive advantage, and exclusive control over critical nodes. We can already see these differences today in how quantum science is being developed. In the United States, progress is driven mostly by large private companies such as Google and IBM, creating a future in which corporate actors could compete, and possibly fight, for control of essential quantum resources. China, by contrast, pursues a state-led model in which quantum networks are closely tied to national security and centralised state authority. Europe follows a third path: the main initiative to build a large-scale quantum network, the Quantum Internet Alliance, is a public-private partnership designed to serve broad public interests. These models also imply different allocations of authority and responsibility. Jasanoff’s (2020) insight is that imaginaries are not neutral backdrops: they shape which futures seem legitimate, which institutions get built, and what forms of authority become normal and acceptable.

Jasanoff's (2020) framework therefore provides a way to examine how competing visions of a future quantum internet may influence its institutional development. The extent to which policymakers approach quantum networking as a shared infrastructure for secure communication or as a strategic resource associated with geopolitical advantage may shape decisions about investment, access, ownership, and control. The relevance of these imaginaries to a quantum governance lens lies in how they interact with the particular technical capabilities and infrastructural requirements of quantum networking, rather than in the assumption that quantum technologies introduce fundamentally new principles of governance.

\subsection{Interscalar vehicles and temporalities}
Hecht (2018) introduces the concept of interscalar vehicles to describe how certain technologies operate as “vehicles that connect phenomena and actors across scales, from the molecular to the planetary” (p. 13). These vehicles do not merely transmit information across scales unchanged, but rather reshape how concepts like responsibility and authority are imagined and assigned. Nuclear science, for example, links atomic reactions to global security regimes, producing what Hecht calls “entangled scales of responsibility” (p. 19). She also shows that technologies bring with them distinct temporalities: the half-life of radioactive waste, the decades-long building of infrastructures, and the near-instantaneous decisions of crisis response all coexist, creating governance challenges that cannot be addressed at a single temporal rhythm. This lesson is particularly apt, since nuclear science itself was born of the first quantum revolution started 100 years ago. Interscalar vehicles thus expose that neither authority nor responsibility can be fixed at one level or moment; they must be continually negotiated in relation to the scales and times a technology binds together.

A global quantum internet exhibits precisely this interscalar and multitemporal character. At one extreme it depends on the fragile coherence of entangled qubits, whose correlations arise and collapse (without protection or error correction) on sub-second timescales. At the other, it relies on global networks of optical fibre, satellites, quantum processing and memory centers, and quantum repeaters, infrastructures that take years to construct and maintain, and that are embedded in the legal and political arrangements of states and international institutions. Its temporalities are likewise layered: the instantaneous non-local correlations of entanglement; the medium-term challenge of ensuring network reliability under fluctuating geopolitical and market conditions; and the long-term implications of embedding non-deterministic processes into critical systems such as finance, health, and defence. This means that decisions about, for example, the allocation of qubit resources such as QRAM or the design of cryptographic protocols have effects that propagate unevenly across local, national, and global levels, and across radically different time horizons.

For governance, these interscalar and temporal dynamics imply that authority and responsibility cannot be stabilised within any single jurisdiction, institution, or time frame. Rather, they must be continually renegotiated as actors move between the technical, organisational, and political scales of the network. A vulnerability discovered in a superposedly-secure protocol (quantum or classical) can immediately raise questions of liability for international data flows; a shift in market access for computational resources can, over the course of years, reshape the global balance of power in AI development. Governance therefore requires modes of coordination and accountability that can operate across scales and adapt to different temporal rhythms: an approach far removed from the static, hierarchical models inherited from classical infrastructure regimes.

\subsection{Accountability, learning, and legitimacy}
’t Hart’s work on crisis governance (Boin, McConnell \& 't Hart, 2008) highlights that legitimacy is not sustained through claims of certainty but through the ability to be accountable and to learn under conditions of disruption. As they put it: “crises are often defined by the loss of meaning, the collapse of routines, and the challenge they pose to established institutions” (p. 5). In such moments, “leaders are judged not by whether they prevent all harm, often impossible, but by whether they act accountably, explain their decisions, and demonstrate a capacity to adapt” (p. 12). They describe crisis governance as having a dual temporal orientation: it must “look backwards to assign responsibility and blame” while also “looking forwards to redesign institutions and prevent future harm” (p. 15). This approach reframes legitimacy as something earned not through control or prediction but through responsiveness and learning. Klein’s (2007) account of disaster capitalism illustrates how crises can be used to justify the transfer of public authority to private actors. In the aftermath of Hurricane Katrina, city elites framed the destruction not as a tragedy to be remedied but as an opening for structural overhaul: “I think we have a clean sheet to start again. And with that clean sheet we have some very big opportunities” (p. 4). Elsewhere, the devastation of public housing was celebrated outright: “We finally cleaned up public housing in New Orleans. We couldn’t do it, but God did” (p. 4). These remarks exemplify how crises can be weaponised to accelerate privatisation and sidestep democratic accountability. Read alongside Boin et al, Klein's analysis raises the possibility that disruptions to emerging quantum infrastructures, or even the anticipation of such disruptions, could similarly be invoked to justify the concentration or transfer of infrastructural control, with consequences for legitimacy and public oversight. For a discussion of potential risks associated with quantum infrastructure disruption and collapse, see Cho (2026).

This dual orientation is relevant to the governance of a quantum internet. As quantum networks develop, disruptions to entanglement distribution, failures in specialised network components, or vulnerabilities in the implementation of quantum communication protocols could affect services that depend on these infrastructures. The consequences would depend on the extent to which quantum networks become integrated into existing communication, financial, and other critical systems. Like classical digital technologies, quantum technologies also have civilian and military applications, raising questions about competing security priorities and the responsibilities of public and private actors. Their specific technical requirements may nevertheless introduce additional operational dependencies. The distribution and preservation of quantum information involve challenges such as decoherence, transmission losses, and the need for error correction, while establishing entanglement between distant nodes requires specialised network components. These challenges do not necessarily make quantum infrastructures less reliable than classical ones, nor do they imply that their failures will be more consequential. Rather, they highlight the importance of understanding how technical dependencies, institutional arrangements, and the growing reliance on shared quantum resources may influence the consequences of disruption and the capacity of institutions to respond.

The difference here is therefore not necessarily one of kind or intensity, but of the particular capabilities and dependencies introduced by quantum technologies. The classical internet has already demonstrated that governance cannot rely on static regulatory frameworks alone, as technological development continually changes the relationships between infrastructure providers, public institutions, and users. Quantum networking may introduce further questions about how authority and responsibility are distributed across specialised and interconnected systems. At the same time, the development of quantum technologies offers opportunities for secure communication, distributed computation, and technical verification, demonstrating that their governance cannot be understood solely through the risks of infrastructural failure. These capabilities may change not only what institutions must protect against, but also how trust can be established and how responsibilities are allocated between those who develop, operate, and use quantum infrastructures. Crisis governance theory helps examine what happens when these arrangements are challenged. As Boin et al. (2008) argue, crises unsettle established expectations and require institutions to make decisions under conditions of incomplete or contested information. In such situations, legitimacy depends not on perfect control but on whether institutions can account for their choices, respond to changing circumstances, and learn from failures.

Quantum infrastructures do not recreate crisis conditions continuously simply because their underlying physical processes are probabilistic. Nevertheless, their development may generate new forms of technical and institutional dependence, while their integration into existing systems could create circumstances in which failures or security concerns require coordinated responses across multiple actors. Such situations raise questions about who has the authority to intervene, who bears responsibility for the consequences of disruption, and whose interests shape the institutional responses that follow. They also bring Klein’s (2007) concerns into view: decisions made in response to actual or anticipated vulnerabilities may have lasting consequences for the distribution of infrastructural control. The relevance of crisis governance to a quantum governance lens therefore lies in examining how authority and responsibility are negotiated when established arrangements are challenged, and how institutions can maintain legitimacy through accountability, responsiveness, and learning. Rather than treating quantum technologies as inherently crisis-prone, this perspective recognises that their development involves both opportunities and risks, and that governance must remain attentive to how the relationships between technological capabilities, institutional authority, and public accountability evolve.

\subsection{Synthesis}
Taken together, these literatures provide a basis for examining a quantum internet as a sociotechnical infrastructure whose development will be shaped by existing governance arrangements, while potentially introducing new capabilities and dependencies that affect how authority, responsibility, and legitimacy are organised. Rohde et al. (2025) highlight not only the technological promise of quantum communication, computation, and sensing, but also the political and economic stakes of allocating and taxing qubit resources, governing transnational networks, and addressing the technical and security challenges associated with quantum communication and computation. Scott’s (1998) critique of high-modernist legibility draws attention to the limitations of governing complex systems through simplified administrative categories, raising questions about which aspects of quantum infrastructures become visible to regulators and which technical and institutional dependencies may remain obscured. Edwards (2003, 2019) shows that infrastructures themselves are architectures of power that organise authority, discipline behaviour, and shape legitimacy, insights underscored by Zuboff’s (2019) demonstration that digital infrastructures of surveillance capitalism extract and exploit behavioural surplus. In this light, applications such as BQC illustrate how the technical design of quantum infrastructures may influence relationships between users and service providers. While particular protocols can provide guarantees concerning the privacy of delegated computations, the institutional arrangements through which computational resources are made available continue to raise questions about access, dependence, and control. Quantum infrastructures may therefore create opportunities to address certain existing asymmetries of information without necessarily resolving broader inequalities in infrastructural power.

Jasanoff’s (2020) analysis of sociotechnical imaginaries draws attention to how competing visions of quantum networking, including its development as a shared communication infrastructure or as a resource for geopolitical advantage, may influence present-day investment priorities and institutional choices. Hecht’s (2018) concept of interscalar vehicles provides a means of examining how particular forms of evidence, technical practices, and institutional processes connect phenomena across spatial, temporal, and organisational scales, making relationships between technical operations and political responsibility visible. Although such relationships are not unique to quantum technologies, this perspective is relevant to understanding how the technical requirements of quantum networking may create dependencies between infrastructure operators, regulatory authorities, and users across different jurisdictions. Boin et al.’s (2008) work on crisis governance further highlights how these institutional relationships may be challenged when disruptions occur, and how legitimacy depends on the capacity of institutions to remain accountable and learn from failure. For example, a disruption affecting a transnational entanglement-distribution network could raise questions about the responsibilities of different operators and regulatory authorities, particularly where the causes and consequences of failure extend across organisational or jurisdictional boundaries. Whether such an event results in greater transparency, institutional learning, or a concentration of authority would depend on existing governance arrangements and the political decisions taken in response.

Together, these perspectives suggest that the governance of a quantum internet cannot be understood solely as the external regulation of a neutral technical system. Formal oversight remains important, but its effectiveness will also depend on how technical capabilities, infrastructural dependencies, and relationships between public and private actors shape the exercise of authority and the allocation of responsibility. These perspectives can be brought together into three interconnected dimensions of a quantum governance lens: the imaginaries shaping infrastructural development, the relationships through which authority and responsibility operate across scales, and the opportunities and constraints that technical design creates for privacy, trust, and power.

First, governance is influenced by the imaginaries that guide technological development (see Section 2.4). Whether quantum networking is approached primarily as a shared infrastructure for secure communication, a commercial opportunity, or a strategic resource may shape decisions about investment, standardisation, ownership, and access. These visions need not be mutually exclusive, and their influence will depend on the institutions and actors involved. Examining them nevertheless helps explain how particular governance arrangements may emerge before quantum networks become widely operational.

Second, governance involves relationships across spatial, temporal, and institutional scales (see Section 2.5). Hecht’s (2018) work provides a way to examine how technical processes and their consequences become connected to questions of responsibility and oversight at different levels. For instance, a disruption to a shared entanglement-distribution service could affect organisations operating across several jurisdictions, raising questions about the responsibilities of infrastructure providers, network operators, and regulatory authorities. Although questions of transnational responsibility are familiar from existing digital infrastructures, the specialised requirements of entanglement generation, distribution, and maintenance may shape which actors possess the technical capacity to identify failures, restore services, and account for disruptions. Examining these dependencies can therefore help establish how responsibility is distributed within particular quantum network architectures.

Third, governance is shaped by infrastructural opportunities and constraints concerning privacy, trust, and power (see Section 2.6). On the one hand, BQC can enable users to delegate quantum computations while concealing aspects of their inputs and computations from the server, subject to the guarantees and assumptions of the particular protocol. This offers possibilities for limiting infrastructure providers’ visibility into users’ computational activity, addressing some of the concerns about asymmetrical access to information raised by Zuboff (2019). On the other hand, these privacy guarantees do not resolve questions about who controls access to quantum computing resources, determines their pricing, or maintains the infrastructure on which users depend. Consequently, the protection of computational privacy may coexist with considerable concentrations of infrastructural power. 

Governance in this context is therefore not about eliminating uncertainty or developing an entirely new set of political principles, but about examining how existing questions of legitimacy, accountability, and equity take shape within emerging quantum infrastructures. A quantum governance lens brings together established perspectives on sociotechnical systems to investigate how the particular capabilities and dependencies of quantum networking may influence institutional arrangements, the distribution of authority, and the possibilities for public oversight. Across these dimensions, questions of accountability and institutional learning remain central, particularly where technological development or infrastructural disruption challenges established arrangements of responsibility and oversight. Its contribution lies not in assuming that quantum technologies fundamentally transform governance, but in identifying where their development may alter the conditions under which familiar governance challenges arise and must be addressed.

\section{A lens for quantum governance}
In the discussion so far we have endeavoured to highlight how a global quantum internet will not simply be a technical innovation but a sociotechnical infrastructure with profound governance implications. To grasp these implications, governance must be approached through a different lens. Rather than treating governance as a fixed structure tasked with managing technological systems, we propose a quantum governance lens that conceives governance as a contextual, adaptive process.

This lens begins from a post-empiricist stance. Data, modelling, and prediction are important resources for governance, but they cannot resolve every question of institutional responsibility, legitimacy, or political judgement. As Edwards (2019) and Scott (1998) illustrate in different ways, the production and organisation of knowledge are themselves shaped by infrastructures, institutional priorities, and practices of simplification. Better information may reduce uncertainty, but it does not eliminate the need to make decisions about competing interests, acceptable risks, and the distribution of responsibility. These considerations are particularly relevant to emerging quantum infrastructures, whose technical capabilities and operational requirements may introduce dependencies that existing governance arrangements must learn to accommodate. Quantum information technologies illustrate the importance of distinguishing technological capabilities from their institutional consequences. Quantum computation and communication exploit physical phenomena such as superposition and entanglement, potentially enabling tasks that are difficult or impractical using classical methods (Nielsen \& Chuang, 2010; Biamonte et al., 2017). However, the practical significance of these capabilities depends on technological development, error correction, infrastructure, and the applications for which they prove useful. As Preskill (2018) emphasises, limitations in the reliability and scalability of quantum devices remain important considerations. The governance implications of quantum technologies should therefore not be inferred from their theoretical capabilities alone, nor should their widespread adoption be treated as inevitable. Rather, a quantum governance lens should examine how emerging technical capabilities interact with institutional choices, infrastructural dependencies, and existing arrangements of authority and responsibility.

Taken together, these observations allow us to define what we mean by a quantum governance lens. Rather than offering a fixed institutional blueprint or a set of requirements dictated by quantum mechanics, we propose a meta-framework for examining how the particular capabilities and infrastructural dependencies of quantum technologies interact with established questions of authority, responsibility, legitimacy, and power. The lens is organised around six analytical and normative commitments:
\begin{enumerate}
    \item \textbf{Recognise the limits of prediction and control} by distinguishing uncertainties associated with technological development, infrastructural operation, and institutional decision-making. Quantum technologies introduce particular technical challenges, but these do not imply that uncertainty is unique to quantum infrastructures or that improved information and modelling cannot reduce it.
    \item \textbf{Examine authority as relational and institutionally mediated} rather than assuming no actor can exercise unilateral control over distributed quantum resources; identifying the actors accountable for the design, operation, maintenance, and oversight of quantum infrastructures. Attention should be paid to how technical dependencies and organisational relationships may complicate the allocation of responsibility without treating technical systems themselves as accountable agents.
    \item \textbf{Examine how responsibility is allocated among human institutions and through technical systems} instead of assigning moral responsibility to hardware or deriving distributed responsibility from entanglement.
    \item \textbf{Emphasise accountability, responsiveness, and institutional learning}  ealongside prediction and prevention, rather than treating these as alternatives.
    \item \textbf{Evaluate privacy-by-design as a normative governance objective} examining how quantum protocols, including BQC and verifiable delegation, may protect users' information while recognising the continuing importance of institutional safeguards, transparency, and accountability.
    \item \textbf{Examine entanglement generation and distribution as resources} whose provision, allocation, maintenance, and control may have governance implications.
\end{enumerate}
These six commitments form the foundation of a quantum governance lens: a way of approaching governance that is structurally compatible with quantum properties such as indeterminacy, non-separability, fragility, and interdependence.

From this perspective, governance can be examined as an ongoing negotiation of legitimacy, authority, and responsibility. Formal political institutions remain important, but their capacity to govern emerging technologies also depends on relationships with infrastructure operators, technical experts, service providers, and users. This relational understanding of authority resonates with aspects of the anarchist tradition, particularly its attention to cooperation and forms of social organisation that do not depend exclusively on centralised command. Proudhon (1840/1994) challenged established forms of political and economic authority, while Kropotkin (1902/2009) emphasised the role of mutual aid in social organisation. These perspectives do not imply that quantum infrastructures are inherently decentralised or that their governance should dispense with public institutions. Rather, they invite examination of how coordination, dependence, and cooperation operate alongside formal structures of authority. 

Responsibility, too, must be examined across the institutions and actors involved in developing and operating quantum infrastructures, including states, corporations, engineers designing error-correcting codes, and cloud providers maintaining access to quantum resources. Although technical systems and infrastructural dependencies shape what these actors can do, responsibility for their design, operation, and consequences remains with the human actors and institutions responsible for them.

Therefore, the quantum governance lens shifts emphasis from prediction to responsiveness. Rather than promising perfect foresight, governance must cultivate transparency understood not as total visibility but as a willingness to remain accountable for decisions made under uncertainty. Edwards’ (2003) account of infrastructures as “the hidden architecture of modernity” (p. 186), Jasanoff’s (2020) insistence on the power of imaginaries in shaping both technologies and institutions and Hecht’s (2018) focus on interscalar dynamics all support this view: governance is not external to infrastructures but emerges through them. The relevance of these perspectives to quantum governance does not lie in the assumption that quantum mechanics introduces fundamentally new principles of political organisation. Rather, it lies in examining how established governance challenges manifest within infrastructures that depend on particular quantum capabilities and technical requirements. The generation and distribution of entanglement, the provision of specialised computational resources, and the development of protocols for secure communication may shape relationships between infrastructure providers, users, and regulatory institutions. These developments raise questions about how responsibility is allocated, how decisions are justified, and how governance arrangements can adapt as technologies mature. Decoupling good governance from the promise of total control therefore remains important, not because quantum technologies uniquely demonstrate the limits of prediction, but because their development provides another context in which accountability, adaptability, and responsiveness must be considered alongside technical performance.

Two further aspects deserve explicit recognition. First, privacy should be considered an important normative dimension of quantum infrastructure governance. As Zuboff (2019) demonstrates in the context of surveillance capitalism, asymmetries in access to information can contribute to unequal relationships of power between infrastructure providers and users. The development of quantum computing services raises related questions about what information providers can access, how that information may be used, and what protections are available to users. Technologies such as blind quantum computation (BQC) offer possibilities for addressing some of these concerns by enabling delegated computation while concealing specified information from the server, subject to the assumptions and guarantees of the relevant protocol. Verifiable variants may additionally allow users to check aspects of the computation's correctness. These capabilities do not eliminate the need for institutional safeguards, nor do they resolve wider questions of access and infrastructural control. Nevertheless, they demonstrate that privacy protections can be incorporated into technical design rather than addressed solely through subsequent regulation. A quantum governance lens should therefore examine how technical privacy guarantees and institutional arrangements can work together to protect users and limit asymmetries of informational power.

Second, entanglement should be recognised as a physical resource whose generation, distribution, and maintenance depend on technical infrastructures and institutional arrangements. Unlike the transmission of classical information over an established connection, the distribution of entanglement requires the preparation and preservation of quantum correlations between systems, subject to challenges such as transmission loss, noise, and decoherence. Long-distance quantum networking may therefore depend on specialised equipment, including quantum repeaters, and on coordination between the actors responsible for operating and maintaining these resources. Although infrastructural dependencies and vulnerabilities are not unique to quantum technologies, the particular requirements of entanglement distribution raise questions about how access, reliability, and responsibility are organised within quantum networks. Disruptions affecting particular links or nodes may compromise services that depend on them, although the wider consequences will depend on network architecture, redundancy, and available alternatives. Similarly, where the provision of entanglement-distribution services becomes concentrated among a limited number of operators, control over these resources may create forms of institutional or economic dependence. A quantum governance lens should therefore examine how decisions about the development, allocation, maintenance, and control of entanglement-related resources shape access, accountability, and the distribution of responsibility, without assuming that their technical properties necessarily determine particular governance outcomes.

In sum, we propose a quantum governance lens that brings established insights from governance theory and science and technology studies into dialogue with the particular capabilities and infrastructural requirements of quantum information technologies. Rather than assuming that quantum mechanics necessitates new political principles or fundamentally transforms existing institutions, this lens examines how emerging quantum infrastructures may affect the relationships through which authority, responsibility, and legitimacy are organised. Its analytical contribution lies in identifying where technical capabilities, infrastructural dependencies, and institutional choices intersect, while its normative commitments emphasise accountability, responsiveness, privacy, and equitable access. In this way, the lens provides a framework for examining the governance implications of quantum technologies without presupposing either their widespread adoption or their eventual institutional consequences.

\section{Applications and illustrations}
Any quantum governance lens can be evaluated by considering how governance questions arise in relation to specific applications of a quantum internet. Rohde et al. (2025) discuss several domains where quantum technologies are expected to deliver transformative capabilities. These examples give evidence towards how a quantum internet can force us to reckon with  authority, responsibility, and legitimacy not being external add-ons to technical systems but rather properties that emerge dynamically through their infrastructures.

\subsection{Quantum computation}
Quantum computation is often regarded as the flagship application of a quantum internet, both technically and politically. As recent analyses show, global investment reflects this emphasis: private funding in quantum technologies is “heavily concentrated in quantum computing,” driven by high-profile breakthroughs such as IBM Q System One and Google’s quantum processors (Devadas \& T., 2025). Rohde et al. (2025) highlight that a fully networked quantum infrastructure would make it possible to link distributed quantum processors, enabling forms of cloud-based computation with the potential to go far beyond today’s classical cloud computation frameworks. They emphasise that “distributed quantum computing will allow us to harness the power of geographically separated devices as if they were a single coherent machine” (p. 243), making computational resources accessible to users without the need to own a large quantum computer or, in architectures using photonic clients, without requiring any quantum hardware on the user side at all.

Rohde et al. (2025) speculate on the emergence of qubit marketplaces, where computational resources are traded, taxed, and leveraged for economic or political advantage. They envision “a marketplace for tradable units of quantum computation, where the underlying asset is time-shares in the global network … subject to ordinary supply and demand” (p. 259), and propose that “the qubit would be ripe for the application of a consumer-driven tax” (p. 271). These proposals illustrate that the allocation of computational resources - who gets to use them, at what price, and under what conditions - cannot be separated from broader questions of governance.

The governance stakes of such markets resonate with the dynamics of surveillance capitalism. As Zuboff (2019) argues, digital platforms have turned human experience into a source of “behavioural surplus” (pp. 8-9), harvested and monetised for predictive and manipulative purposes. In a quantum-enabled economy, the power to control and monetise access to scarce computational resources risks creating a new layer of behavioural surplus: data about how and where qubits are used could become a valuable commodity. Moreover, quantum-enhanced AI, building on the computational advantages of quantum hardware (Biamonte et al., 2017, pp. 195-196; Huang, Kueng, \& Preskill, 2021, pp. 2-3), would amplify the capacity of those who control both computation and data to shape economic and political processes. Governance of qubit markets thus requires attention not only to pricing and taxation but also to the infrastructural embedding of authority and responsibility in market design itself.

The capacity to run algorithms in polynomial time instead of exponential time, including for breaking cryptographic encryption systems, has immediate governance significance. If quantum computation becomes essential for tasks such as secure communication, medical research, or climate modelling, then access to quantum resources will determine who can participate meaningfully in critical domains of knowledge and innovation. Rohde et al. (2025) anticipate “a marketplace for tradable units of quantum computation, where the underlying asset is time-shares in the global network … subject to ordinary supply and demand” (p. 259), and even propose that “the qubit would be ripe for the application of a consumer-driven tax” (p. 271). Such proposals draw attention to the relationship between ownership of quantum hardware and control over access to computational resources. Although ownership remains an important source of authority, the development of shared or distributed quantum computing services may also give infrastructure operators and service providers considerable influence over how computational capacity is allocated, priced, and accessed.

From a quantum governance perspective, this shift means that questions of authority, responsibility, and legitimacy are built into the architecture of quantum computation. Decisions about who allocates computational time, who sets pricing and taxation, and who guarantees equitable access are inseparable from the technical protocols that enable distributed computation. The concentration of quantum resources in a few powerful actors, whether states, corporations, or hybrid alliances, risks amplifying existing global inequalities. At the same time, the non-local nature of quantum network applications challenge governance approaches that assume clearly identifiable points of control. Authority cannot be fully separated from the infrastructural processes that configure computational capacity, and responsibility for outcomes will often be distributed across human and non-human actors. 

Ownership remains an important consideration in the governance of quantum computation, but it does not fully capture the relationships through which computational resources are made available and controlled. As with existing cloud computing infrastructures, access may depend on arrangements involving hardware owners, service providers, and infrastructure operators. Quantum computing introduces additional technical requirements, including specialised operating environments and, in distributed architectures, the coordination of quantum processors and entanglement-distribution services. These dependencies warrant examination alongside questions of ownership, licensing, pricing, and regulation.Examining these relationships is particularly important where access to computational capacity depends on a limited number of providers, since decisions about resource allocation, pricing, and service provision may reinforce existing inequalities in access and concentrate infrastructural power.

A quantum governance framework takes a different starting point. Rather than focusing on devices in isolation, it treats access, not ownership, as the principal governable object. Because meaningful quantum computation increasingly depends on collaboration across networked processors, entanglement distribution, and shared datasets, governance must concern itself with the fairness of scheduling, the allocation of network resources, and the conditions under which users participate in distributed computation. In this lens, the objective is not only to regulate who owns a machine, but to ensure that the benefits of quantum computation are not determined solely by market power or the happenstance of physical infrastructure. By emphasising interoperability, equitable access, and the relational character of quantum networks, a quantum governance framework addresses precisely the vulnerabilities that a classical approach would overlook.

\subsection{Secure communication: quantum key distribution}
One of the earliest and most developed applications of quantum network technologies is quantum key distribution (QKD), which enables two parties to exchange encryption keys with the guarantee that any eavesdropping attempt will be detectable. QKD derives its security not from the presumed difficulty of solving certain mathematical problems, as in classical public-key cryptography, but from the laws of quantum physics. As Rohde et al. (2025) explain, “any attempt to eavesdrop on a quantum channel introduces detectable disturbances” (p. 244), making interception physically detectable in both theory (Bennett \& Brassard, 1984), and practice (Storz et al., 2023).

This distinction is crucial in light of the threat that large-scale quantum computation poses to existing encryption systems. Widely used public-key cryptosystems such as RSA rely on the classical intractability of factoring large integers. However, as Nielsen and Chuang (2010) note, “Shor’s factoring algorithm runs in polynomial time on a quantum computer, thereby undermining the security of widely used public-key cryptosystems” (pp. 221-222). The development of powerful quantum computers would thus render classical encryption schemes vulnerable, creating a pressing need for alternatives whose security does not depend on assumptions about computational hardness.

QKD meets this need by providing security guaranteed by the laws of physics. From a governance perspective, however, this does not eliminate questions of power. The ability to communicate securely will depend on who controls access to QKD infrastructures and under what conditions. If states monopolise QKD networks, they may consolidate surveillance capacities at end user terminals to gain geopolitical leverage. On the other hand, if corporations dominate, access may become stratified by wealth and market power. Authority and responsibility are therefore embedded not only in the physical principles that enable QKD but also in the infrastructural arrangements, ownership, access protocols, standards, that determine how those principles are realised in practice. Transparency and legitimacy depend less on the intrinsic properties of QKD than on how its infrastructures are governed.

From this perspective, a classical governance framework risks making a decisive category error. By treating encryption as a technical subsystem that can be certified and regulated independently much as classical public-key schemes are validated by standards bodies, it presumes a clean separation between secure communication and the institutional structures that authenticate and supervise users. In a quantum setting, that separation collapses. The effectiveness of QKD depends on who controls the distribution of quantum states, who is authorised to initialise and authenticate communication partners, and how access to the physical network is regulated. If governance focuses only on certifying the protocol while ignoring control over the underlying infrastructure, it may inadvertently enable the very monopolies it seeks to prevent: a single state agency, corporation, or consortium could unilaterally determine who can communicate securely and under what conditions. A quantum governance framework would proceed differently. It would recognise that secure communication and the authority to verify identities are inseparable, and would prioritise preventing any single entity from monopolising the authentication layer. This requires embedding plurality, interoperability, and distributed trust directly into infrastructural design, rather than relying on ex post oversight. Only by treating QKD as both a physical and institutional system can governance avoid reproducing classical mistakes in a domain where communication security and surveillance capacity arise from the same quantum processes.

\subsection{Distributed consensus: Quantum Byzantine Agreement}
Quantum Byzantine Agreement (QBA) extends the study of distributed consensus by examining how quantum communication and other quantum resources may be used to achieve agreement under particular assumptions about faults, adversarial behaviour, and network capabilities. As Rohde et al. (2025) highlight, quantum protocols offer possibilities for coordinating distributed participants, potentially providing advantages over classical approaches in specific settings. These advantages may concern communication requirements, security guarantees, or the conditions under which agreement can be reached, but depend on the assumptions and resources of the particular protocol. QBA should therefore not be understood as universally more resilient or fault-tolerant than classical Byzantine Agreement, but as an approach that introduces additional technical possibilities for addressing established problems of distributed consensus.

The governance implications of QBA concern the relationship between the technical guarantees of consensus protocols and the institutional arrangements through which distributed systems are operated. Classical Byzantine Agreement protocols already demonstrate that agreement can be achieved despite certain forms of failure or malicious behaviour, without necessarily identifying every faulty participant. Quantum variants do not fundamentally change this principle, but may introduce different assumptions concerning quantum communication channels, shared quantum resources, and the capabilities of adversarial participants. These distinctions matter because the reliability of a consensus mechanism depends not only on its theoretical guarantees but also on whether the conditions underlying those guarantees can be maintained in practice.

Depending on the protocol, these conditions may include the availability and quality of quantum communication channels, the preparation and distribution of entangled states, or participants' ability to perform specified quantum operations. Maintaining them may require specialised infrastructure and coordination between network operators, protocol designers, and participating institutions. This raises questions about who is responsible for providing and maintaining the necessary resources, how the reliability of the protocol can be assessed, and how disruptions or malicious behaviour should be addressed. Although such questions are familiar from classical distributed systems, the particular technical requirements of quantum protocols may influence how responsibilities are allocated and exercised.

A quantum governance lens therefore examines how accountability for the behaviour of individual participants interacts with responsibility for the technical infrastructure and institutional arrangements supporting consensus. It does not assume that responsibility emerges from quantum interactions themselves or that governance must abandon the identification of faulty actors. Rather, it directs attention to the relationships between protocol design, infrastructure provision, institutional oversight, and the practical conditions under which agreement can be achieved. This includes examining who establishes the conditions for participation, who can assess whether a protocol is operating as intended, and how responsibilities are negotiated when multiple organisations contribute to its operation.

QBA thus illustrates how emerging quantum capabilities can be examined through established questions of authority, responsibility, and legitimacy. Its contribution to a quantum governance lens lies not in demonstrating that quantum mechanics necessitates a fundamentally different form of governance, but in identifying how the technical assumptions and infrastructural dependencies of particular consensus protocols may influence the organisation of oversight, accountability, and institutional cooperation.

\subsection{Blind Quantum Computation}
BQC is another application highlighted by Rohde et al. (2025) that exemplifies the entanglement of technical capabilities and governance concerns. BQC allows a user to delegate a computation to a remote quantum server without revealing either the data or the algorithm. As Rohde et al. describe it, “the server learns nothing about the data or the algorithm being computed” (p. 244). This promises unprecedented levels of privacy in outsourced computation, making it possible for clients to use powerful remote resources while maintaining control over sensitive information. Crucially, recent advances in verifiable BQC(e.g., Leichtle et al., 2021; Kashefi et al., 2024) show that even noisy or partially untrusted devices can be used in ways that allow the client to verify that the computation was performed correctly without revealing anything about its content. Experimental demonstrations (e.g., Barz et al., 2012) further illustrate that clients may only need to prepare and transmit simple photonic states rather than operate a full quantum computer.

From a governance perspective, however, the deployment of BQC raises complex issues. On the one hand, it subverts the behavioural surplus described by Zuboff (2019), since it prevents servers from harvesting user data as a by-product of computational services. Servers cannot infer what is being computed, on which data, or with what purpose: a foundational rupture from classical cloud computing, where data extraction is a structural feature rather than a bug. This calls into question the economic models underpinning platform capitalism, which rely precisely on the asymmetry between users who reveal information and firms that extract, correlate, and monetise it. 

On the other hand, BQC complicates oversight and accountability. If the computation is hidden from the server, and if server and client are located in different countries, who is responsible for ensuring that the computation does not violate regulation or law? How can regulators oversee the training of AI models if the model, the data, and even the training procedure may be concealed from any external observer? Blindness may protect individuals and small organisations, but it also risks enabling activities (such as privacy-violating AI, financial manipulation, or prohibited research) that are impossible to detect if enforced through classical visibility-based governance frameworks. The governance challenge is therefore not simply to protect privacy but to ensure that the shield of blindness does not become an avenue for unaccountable power.

To see the stakes clearly, we must make explicit the distinction between what a classical governance framework assumes, where it fails, and what a quantum governance framework must do instead.

In classical computing, secrecy is treated as a technical feature: an optional privacy service layered on top of computational infrastructure. Governance therefore rests on oversight through the capacity for transparency. Regulators assume that, even if computation is encrypted, it can ultimately be made legible through auditability requirements, reproducibility of computations, access to logs, and trusted intermediaries capable of inspecting the process. Security and secrecy are thus balanced against the principle that regulators can, when necessary, “look inside” the computation, or at least obtain a traceable explanation of how it was performed. This assumption underpins regulatory approaches ranging from financial audits to medical-data governance to the oversight of classical AI systems.

The classical assumption fails catastrophically in the quantum case because blindness is not optional and not reversible. A blind quantum computation cannot be “opened up” for inspection without destroying the quantum states involved. Any attempt to force auditability,  for example, by requiring reproducible outputs or server-side process transparency, breaks blindness and thereby reintroduces the very privacy vulnerabilities BQC is designed to eliminate. In other words: requiring reproducibility or inspectability makes blind computation impossible; it forces users to reveal sensitive information to servers they cannot reasonably trust. 

The mistake is to treat blindness as a technical privacy layer rather than a fundamental property of the computation itself. Classical governance frameworks tend to prioritise visibility and auditability. When applied to BQC, this creates tension: regulatory mechanisms designed around inspectability may inadvertently undermine the very privacy guarantees that make BQC valuable. This becomes even more apparent with blind quantum machine learning (Fleur, 2025) and its near term implementations (Pasini et al., 2026). When models can be trained blindly, regulators cannot rely on inspecting training data, architectures, or intermediate states. Regulatory approaches that depend on direct inspection of computational processes may therefore encounter difficulties when applied to blind quantum machine learning. This raises questions about how compliance can be assessed without unnecessarily compromising the privacy guarantees that make such technologies valuable.

A quantum governance framework must therefore shift from oversight through visibility to oversight through verifiability. Instead of asking: “Can regulators see what the computation is?” the relevant question becomes: “Can regulators verify that the computation meets certain protocol-level guarantees, without seeing any private information?” This becomes a problem of certification, which BQC can also be used to do (Broadbent, 2009).

This is the heart of the shift enabled by BQC:
\begin{itemize}
    \item Verification replaces inspection: Regulators need only check that the computational protocol is sound, not what the computation contains.
    \item Correctness is guaranteed statistically, not through reproducibility: Verification schemes allow clients or third-party auditors to confirm that the computation was executed faithfully without observing the data or the algorithm.
    \item Compliance becomes structural: A quantum governance lens evaluates whether the \textit{infrastructure itself} enforces compliance (e.g. through blind-verifiable protocols) rather than relying on after-the-fact audits.
    \item Responsibility is assigned at the protocol level: Because servers cannot know the content of the computation, responsibility for misuse shifts towards the clients, protocol designers, and the legal frameworks governing allowable uses rather than the holder of the physical server. This reframing is essential; if regulators insist on classical visibility-based oversight, blind quantum machine learning and blind delegated computation will be forced into illegality or will be implemented in weakened, insecure forms. If regulators instead adopt protocol-level verification and normative constraints embedded into infrastructure design, BQC becomes not a threat but a tool for protecting privacy, resisting surveillance capitalism, and enabling AI development that does not rely on mass data extraction.
\end{itemize}

In short, the governance mistake is to attempt to “look inside” computations that are physically structured to resist inspection. The quantum governance alternative is to build regulatory legitimacy on verifiability without visibility, aligning governance with the physical constraints of quantum information rather than fighting against them.

\subsection{Quantum sensing}
Quantum sensing - using quantum phenomena such as superposition and entanglement to achieve ultra-precise measurements - offers capabilities with both beneficial and disruptive potential. Rohde et al. (2025) note applications ranging from navigation and geophysical exploration to medical diagnostics and environmental monitoring. These technologies promise to enhance scientific knowledge and public services, but they also have significant security and political implications. For example, quantum sensors could detect hidden submarines or underground installations, altering strategic balances and creating new vulnerabilities.

The governance implications of quantum sensing stem from its dual-use character and its dependence on distributed infrastructural networks. Decisions about who has access to sensing data, how it is stored and shared, and under what legal and ethical constraints it can be used will shape its effects on both domestic and international security. As with other quantum applications, authority and responsibility here are embedded in infrastructural choices: in the design of data-sharing protocols, the oversight of sensor deployment, and the standards that regulate their use. Moreover, recent work on private and robust states for distributed quantum sensing demonstrates that privacy can be built into the sensing process itself, ensuring that only information about a global function of parameters, and not local, sensitive information, is revealed to participants (Bugalho et al., 2025). This shows that privacy in sensing is not a post-hoc condition but can be an intrinsic property of the quantum states used.

A classical governance framework would treat sensing as a passive operation performed on an already existing world. Data is presumed to pre-exist “out there” and governance focuses on regulating storage, access, and use after the fact. This is a mistake in the context of quantum sensing. Because quantum sensors interact coherently with physical systems, the act of sensing itself changes what can be known and by whom. The increased capacity for high-resolution, non-local, and potentially covert measurement arises at the physical layer. By the time data is recorded classically, it may already have produced irreversible strategic, commercial, or political consequences. Moreover, because quantum sensing can reveal aggregate functions of distributed parameters without revealing the underlying local values (as shown in Bugalho et al., 2025), governance approaches premised on auditability of raw data will fail: there may be no raw data to audit in the conventional sense.

A quantum governance framework instead treats sensing as an active and collaborative process. It recognises that quantum sensing can be combined with privacy-preserving techniques and therefore insists that the design of sensing infrastructures must embed such guarantees from the outset. Rather than relying on the visibility of collected data, governance focuses on protocol-level verifiability, ensuring that authorised users can confirm the integrity and legitimacy of the sensing process without requiring access to the full underlying measurements. This reframes sensing as something that people participate in, through shared entangled states, distributed resources, and negotiated access conditions, rather than something done to them. It also allows governance to intervene at the right level: not after quantum data has already been extracted, but at the level of how the physical interactions that produce that data are arranged, shared, and constrained.

\subsection{Implications}
The applications examined in this section demonstrate that the development of a quantum internet presents opportunities that extend beyond improvements in technical performance. Distributed quantum computation could make specialised computational resources accessible to users who do not own quantum hardware. Quantum key distribution may provide additional means of securing communications under particular technical and operational conditions, while quantum consensus protocols could support coordination between distributed participants. Blind quantum computation offers possibilities for protecting information during delegated computation, and quantum sensing may enable forms of measurement that are difficult to achieve using classical technologies. Although these applications differ considerably in their technical requirements and maturity, they illustrate the potential benefits of quantum information technologies and the importance of examining how those benefits might be realised and distributed.

The governance implications of these developments cannot be understood solely by considering the individual technologies in isolation. Their practical operation will depend on relationships between hardware providers, network operators, protocol designers, regulatory authorities, and users. These relationships are familiar from existing digital infrastructures, but particular quantum capabilities and requirements may introduce additional dependencies or alter the significance of existing ones. A quantum governance lens therefore directs attention to how technical design and institutional arrangements interact, without assuming that quantum technologies necessarily require fundamentally new principles of political organisation.

Across the applications discussed, several considerations emerge. In distributed quantum computation, questions of ownership must be considered alongside the allocation, pricing, and accessibility of computational resources. For quantum key distribution, the security guarantees of particular protocols must be examined together with authentication requirements, implementation choices, and the institutional arrangements governing access to communication infrastructure. Quantum Byzantine Agreement raises questions about the assumptions under which distributed participants can reach agreement, including the capabilities and responsibilities of the actors operating the network. Blind quantum computation illustrates how privacy protections can be incorporated into technical protocols, while also demonstrating that computational privacy does not necessarily eliminate dependence on infrastructure providers. Quantum sensing, meanwhile, invites consideration of how new measurement capabilities may affect access to information, surveillance practices, and the distribution of knowledge and power. These examples do not establish that existing governance approaches are inadequate in every case. Rather, they identify where particular technical characteristics warrant closer attention when evaluating institutional arrangements.

The development of a quantum internet also presents choices about the organisation of shared infrastructure. Decisions concerning interoperability, technical standards, ownership, access conditions, and the provision of specialised resources may influence which actors can participate in emerging quantum networks and on what terms. Where essential services become concentrated among a limited number of providers, questions of dependence, accountability, and equitable access may become especially important. Conversely, arrangements that support cooperation between institutions, transparent allocation procedures, and appropriate safeguards for users may help broaden participation and distribute the benefits of technological development. These outcomes will not be determined by quantum mechanics alone, but by the interaction between technical possibilities, economic incentives, and political and institutional decisions.

A quantum governance lens offers a means of evaluating these choices while recognising both the opportunities and the uncertainties associated with emerging technologies. Its purpose is not to replace existing governance mechanisms with a fundamentally different political order, but to examine whether established arrangements adequately address the capabilities and dependencies of particular quantum applications. This includes identifying the actors responsible for technical and institutional decisions, assessing how risks and benefits are distributed, and considering how accountability can be maintained as infrastructures develop. Prediction, regulation, and formal oversight remain essential, but they must be complemented by responsiveness, institutional learning, and an awareness of how technical design can enable or constrain the exercise of authority.

The potential significance of a quantum internet therefore lies not only in what its technologies may eventually make possible, but also in the institutional choices through which those possibilities are developed. Secure communication, distributed computation, privacy-preserving services, and new measurement capabilities could provide considerable benefits, although their scale and practical significance remain uncertain. Whether these benefits become widely accessible, reinforce existing concentrations of power, or produce new forms of dependence will depend in part on decisions made during the development of quantum infrastructures. By bringing questions of authority, responsibility, legitimacy, privacy, and equitable access into the examination of technical design and infrastructural organisation, a quantum governance lens provides a way to investigate these decisions before particular arrangements become firmly established.

\section{Conclusions and future work}
The second quantum revolution will not only advance society, but challenge the structure of governance at a foundational level, as we have illustrated in our cases studies. In particular, a quantum internet has the potential to reshape how we communicate, compute, and coordinate in ways comparable to the rise of the classical internet. Yet treating its governance as an extension of classical oversight is insufficient. The quantum internet is not a neutral tool waiting to be regulated; it is a sociotechnical system in which authority, responsibility, and legitimacy can emerge through the design and operation of infrastructures themselves. 

Classical governance frameworks assume that uncertainty can be controlled or eliminated. Scott’s (1998) critique of high-modernist schemes shows why this is risky: such approaches become brittle when the system does not behave as expected. Quantum infrastructures bring this point into sharp focus. Entanglement, non-determinism, fragility, and interdependence are structural features, not temporary challenges that better engineering will remove. Treating infrastructures as external to governance also misses Edwards’ (2003, 2019) insight that infrastructures are part of political order: they shape authority and legitimacy as much as they support technical operations. Jasanoff’s (2020) account of imaginaries reminds us that the governance of a quantum internet is already being shaped by competing visions of security, cooperation, and rivalry, while Hecht (2018) shows that quantum infrastructures operate across scales and times in ways that complicate simple institutional control. Boin et al's work on accountability and learning (2008) highlights that legitimacy depends not on perfect control but on the willingness to explain decisions, learn from failures, and adapt.

The quantum governance lens developed in this paper responds to these challenges. It shifts attention away from prediction and full visibility, and towards responsiveness, accountability, and the infrastructural conditions under which legitimate decisions can be made. It emphasises that governance does not sit above technological systems; it emerges through them. Authority becomes relational, responsibility becomes distributed, and legitimacy depends on governance practices that work with uncertainty rather than denying it.

Importantly, this is not a story of inevitability; the future is not fixed. Quantum technologies open space for better choices around governance. Quantum infrastructures reveal limits of classical control, and in doing so, they give us an opportunity to build governance systems that are more open, more accountable, and more democratic, even without an underlying quantum infrastructure. The opportunity lies in recognising that adaptability and responsiveness are strengths, not signs of failure. If institutions choose to work with these ideas, legitimacy in the quantum era can rest on the active practice of accountability under uncertainty, not the passive promise of foresight. If these issues are left unaddressed, future governance will have points of failure built in due to an improper lens for understanding and interacting with quantum technologies. 

This has implications well beyond quantum networks. Many emerging technologies, including artificial intelligence,  already share similar qualities of opacity, interdependence, and unpredictability. The quantum governance lens developed here can therefore inform current debates on AI governance. Rather than relying on assumptions that every process must be fully legible or predictable, policymakers can adopt approaches that support contextual responsiveness, verifiability without visibility, distributed responsibility, and privacy-by-design. These principles are compatible with quantum systems, but they also offer constructive ways to govern today’s opaque algorithmic systems before quantum technologies arrive at scale.

A further point is worth emphasising: we do not need to wait for a quantum internet to implement more privacy-preserving and consensus-oriented governance structures. These approaches can be adopted through regulation, institutional design, and democratic decision-making today. Doing so now would not only strengthen current governance systems but also prepare societies for the demands of the second quantum revolution. And if we fail to do so, we risk building quantum infrastructures that reproduce the inequalities, extractive logics, and brittle hierarchies already visible in classical digital systems. The aim is not technological determinism, but a modest claim: quantum technologies will work best for society if our governance choices are already aligned with the values they make possible.

There are also several directions for future work. This paper has focused on communication, computation, consensus, sensing, and markets, but other quantum technologies will raise new questions as they develop. Future analysis could explore how quantum memory, error-corrected long-distance networks, quantum-secure digital identity, or hybrid classical–quantum AI systems will reshape governance. Another important area is how international institutions will coordinate transboundary quantum infrastructures, and how accountability can be maintained when failures or attacks propagate across jurisdictions in ways classical governance is not prepared for. Finally, more interdisciplinary work is needed to map how quantum concepts like context, non-separability, and entanglement might influence governance theory even outside the technological domain. We hope that this preliminary investigation of quantum governance will support others investigating these questions in the future.

\section{Acknowledgements}

 SKHL gratefully acknowledges support from the Research Master in Public Administration and Organisation Science at Utrecht University.
 
 TzBP gratefully acknowledges support from the Quantum Software Consortium Ada Lovelace Fellowship, and funding under European Union's Horizon Europe research and innovation programme under grant agreement No.~101102140 – QIA Phase 1. 

 We also thank Hidde Stoffels for his helpful comments and careful proofreading of the manuscript.
 
\nocite{*}
\addcontentsline{toc}{section}{References}
\printbibliography
\end{document}